# Solving the MIT Inverse Problem by Considering Skin and Proximity Effects in Coils

Hassan Yazdanian, Reza Jafari, *Senior Member, IEEE,* and Hamid Abrishami Moghaddam

*Abstract—* This paper presents an improved technique for solving the inverse problem in magnetic induction tomography (MIT) by considering skin and proximity effects in coils. MIT is a non-contact, noninvasive, and low-cost imaging modality for obtaining the distribution of conductivity inside an object. Reconstruction of low conductivity distribution by MIT requires more accurate techniques since measured signals are inherently weak and the reconstruction problem is highly nonlinear and ill-posed. Previous MIT inverse problem studies have ignored skin and proximity effects inside coils in the forward method. In this article, the improved technique incorporates these effects in the forward method. Furthermore, it employs the regularized Gauss-Newton algorithm to reconstruct the conductivity distribution. The regularization parameter is obtained by an adaptive method using the two input parameters: a coefficient and an initial conductivity distribution. The new Jacobian matrix is computed based on a standard technique. To compare the early and improved forward methods in possible medical and industrial applications with low conductivity regions, a 2D 8-coil MIT system is modeled, and image reconstruction is performed for synthetic phantoms. Results show that it is crucial to use the improved forward method for the reconstruction of the absolute conductivity values.

## I. Introduction

MAGNETIC induction tomography (MIT) is an imaging modality that attempts to reconstruct the conductivity distribution inside an object [1]–[4]. Its non-contact, non-invasive, non-radiative, and low-cost features make this technique extremely attractive for a wide range of industrial and biomedical applications. Liquid metal visualization [5], monitoring of steel solidification [6], multi-phase flow imaging [7], and intracranial hemorrhage imaging [8], [9] are examples of these applications. The first two applications are categorized in the high conductivity MIT with conductivity values greater than $10^6$ S/m and the next two applications are classified in the low conductivity MIT with conductivity values less than 10 S/m [10].

In a typical MIT system, an array of coils is placed on the boundary of imaging space and acts as both the exciter and sensor. The imaging region is exposed to an alternating magnetic field by one coil which is called *primary magnetic field* and eddy currents are induced inside the conductive objects. These eddy currents generate another magnetic field called *secondary magnetic field*. The voltage in the MIT coils is induced by the *total magnetic field* which is the sum of the primary and secondary fields [11]. In the low conductivity MIT, compared to the high conductivity MIT, the secondary magnetic field is generally much weaker than the primary magnetic field because the conductivities to be imaged are considerably lower in this category. Consequently, the induced voltages are weak in the low conductivity MIT and their detection not only requires the improvement of hardware component but also necessitates a more elaborate computational part in order to acquire more reliable imaging data.

So far, a wide variety of approaches have been proposed to improve the image reconstruction in low conductivity MIT. In some studies, difference imaging techniques have been considered. For instance, time [12], dual-frequency [13], and multi-frequency [14] difference imaging techniques have been proposed for improving intracranial hemorrhage imaging. Some studies have been carried out to reduce the computational cost of forward problem in low conductivity applications [15]–[18]. MIT signal denoising using wavelet transform has been applied in [19] to improve the quality and resolution of reconstructed images. In [20], three new nonlinear reconstruction methods have been implemented for absolute imaging of low conductivity material distributions. In order to improve the reconstruction accuracy and achieve fast imaging in low conductivity applications, artificial intelligence algorithms have been employed in some studies [21], [22]. New coil arrangements such as an open cambered MIT [8] and combined planar MIT [9] have been proposed to improve the sensitivity for the detection of intracranial hemorrhage. In [7], the authors performed a theoretical/numerical analysis of single-sensing and differential gradiometer coils. They demonstrated that the latter enhances significantly the sensitivity of the induced signal to the conductivity.

It makes sense to increase the operating frequency in order to strengthen the magnitude of secondary magnetic field and, consequently, improve the quality of reconstructed image. However, increasing the frequency intensifies skin and proximity effects in conductors that are exposed by alternating magnetic fields [23]. It means, unlike static cases, the current densities inside coils are no longer a constant space function

The authors are with the Department of Biomedical Engineering, Electrical Engineering Faculty, K. N. Toosi University of Technology, Tehran, Iran (e-mail: h.yazdanian@ee.kntu.ac.ir; jafari@eetd.kntu.ac.ir; moghaddam@kntu.ac.ir).

and independent of position of coils relative to each other. Yazdanian *et al.* proposed an *improved forward method* by considering the skin and proximity effects inside MIT coils [24]. This method was in contrast to the *early forward method*, which ignored skin and proximity effects inside MIT coils. Their findings showed that applying the improved forward method is substantial to model 2D MIT coils especially at low conductivity applications in which induced voltages are inherently weak. However, the importance of using the improved forward method has not been studied yet for conductivity reconstruction.

In this study, we investigate the impact of considering the skin and proximity effects in MIT coil on the reconstructed conductivity images for the first time. Thus, the first contribution of this paper is to employ the improved forward method [24] in the solution of the inverse problem. To solve the inverse problem, the Gauss-Newton (GN) algorithm [25] is employed. This algorithm required to calculate the Jacobian matrix which is obtain from the forward problem. The second contribution of this paper is to obtain a new Jacobian matrix based on the improved forward method.

The paper is organized as follows: in Section II.A, the MIT forward problem formulation is introduced and equations for the early and improved methods are presented. In Section II.B, the regularized GN algorithm for MIT conductivity image reconstruction and the method for choosing the regularization parameter are presented. In Section II.C, the sensitivity and standard techniques are introduced for computing the Jacobian matrix. Then, in Section III, through modeling a 2D (two dimensional) 8-coil MIT system and using synthetic phantoms, the conductivity image reconstruction based on the improved forward method will be investigated and compared to that of the early one. The results are discussed and analyzed in Section IV. Finally, Section V presents the conclusion.

## II. METHODS

### A. MIT Forward Problem

In the MIT forward problem electromagnetic field equations are solved for a given conductivity distribution and problem geometry when exciter coils are activated with different patterns. The governing equations are obtained under the magneto-quasi-static approximation [26] in which it is assumed the displacement current can be neglected. This assumption is applicable for a wide range of industrial and medical applications of MIT.

#### 1) Early Forward Method

Early forward method uses simplified Maxwell's equations in which a constant and position-independent total current density (TCD) is assumed inside the coils. This simplification is equivalent to ignoring skin effect. Moreover, it is assumed that TCD is independent of relative position of the coils. This assumption is equivalent to ignoring proximity effect [24]. The governing equations for the early method in the frequency domain can be stated as follows [24]:

$$\begin{cases} \dfrac{1}{\mu}\nabla^2 \bar{A}_z = -\bar{J}_k = -\dfrac{\bar{I}_k}{S_k} & \text{in} \quad \Omega_S \\ \dfrac{1}{\mu}\nabla^2 \bar{A}_z - j\omega\sigma \bar{A}_z = 0 & \text{in} \quad \Omega_{SF} \end{cases} \quad (1)$$

where $\bar{A}_z$, $\bar{J}_k$, $\bar{I}_k$, and $S_k$ are the z-component of the magnetic vector potential (MVP), phasor TCD, phasor current, and cross section area of the k-th current-carrying conductor, respectively, and $j=\sqrt{-1}$. Domains $\Omega_S$ and $\Omega_{SF}$ are the source (coil) and source-free regions, respectively. The bar mark indicates a complex-valued variable and $\mu$, $\sigma$, and $\omega$ are magnetic permeability, electrical conductivity, and angular frequency, respectively. Boundary conditions are required to obtain a unique solution for (1). At infinity, all the fields must vanish. At the interface of the domain, the boundary conditions are given by [27]:

$$\bar{A}_z|_{\Omega_{SF}} = \bar{A}_z|_{\Omega_S}$$
$$\dfrac{1}{\mu_{SF}}\nabla \bar{A}_z \cdot \vec{n}|_{\Omega_{SF}} = \dfrac{1}{\mu_S}\nabla \bar{A}_z \cdot \vec{n}|_{\Omega_S} \quad (1a)$$

where $\vec{n}$ is the outward unit normal and $\mu_{SF}$ and $\mu_S$ are the permeability in $\Omega_{SF}$ and $\Omega_S$, respectively. The boundary condition (1a) is valid for the Coulomb gauge only [28]. Boundary Equation (1) together with boundary condition (1a) constitutes the forward problem in 2D MIT based on the early method.

A discretized finite element (FE) equivalent of (1) is given as follows [29]:

$$\left[\dfrac{1}{\mu}S + j\omega\sigma\kappa\mathcal{T}\right]\bar{\mathcal{A}} = \mathcal{T}\bar{\mathcal{J}} \quad (2)$$

where the column matrices $\bar{\mathcal{A}}_{r\times 1}$ and $\bar{\mathcal{J}}_{r\times 1}$ are the phasors of the node potentials and total current density values, respectively, and $r$ is the number of interpolation nodes. The square matrices $S_{r\times r}$ and $\mathcal{T}_{r\times r}$ are the usual FE coefficient matrices. Note that the constant $\kappa$ in (2) is equal to zero and unity for the FE elements in $\Omega_S$ and $\Omega_{SF}$, respectively.

#### 2) Improved Forward Method

In the improved forward problem method, skin and proximity effects in the exciter and sensor coils are incorporated. Consideration of skin and proximity effects requires the use of a position-dependent TCD in Maxwell's equations inside the coil domain [24]. The governing equations for the improved method in the frequency domain can be stated as follows [24]:

$$\begin{cases} \dfrac{1}{\mu}\nabla^2 \bar{A}_z - j\omega\sigma\bar{A}_z \\ \quad + j\omega\dfrac{\sigma}{S_k}\iint_{R_k}\bar{A}_z(x,y)\,\mathrm{d}s = -\dfrac{\bar{I}_k}{S_k} & \text{in} \quad \Omega_S \\ \dfrac{1}{\mu}\nabla^2 \bar{A}_z - j\omega\sigma \bar{A}_z = 0 & \text{in} \quad \Omega_{SF}. \end{cases} \quad (3)$$

Boundary conditions similar to (1a) are imposed to (3). Equation (3) together with boundary conditions constitutes the forward problem formulation in 2D MIT based on the improved method. This formulation considers skin and proximity effects in the source (coil) region.

The discretized FE equivalent of (3) is obtained from the following equation [29]:

$$\left[\frac{1}{\mu}\mathcal{S} + j\omega\sigma(\mathcal{T} - Q\mathcal{P}^{-1}Q^{T})\right]\bar{\mathcal{A}} = Q\mathcal{P}^{-1}\bar{I} \quad (4)$$

where the rectangular matrix $Q$ and diagonal matrix $\mathcal{P}$ are explained in [29].

*3) Induced Voltage*

The output of the forward problem for a one turn sensing coil is the induced voltage as obtained by [24]:

$$\bar{V} = j\omega \oint_C \bar{A}_z \, \vec{a}_z \cdot d\vec{\ell} \quad (5)$$

where $C$ is a closed contour bounding the internal area of the sensing coil. For a 2D coil, only the straight segments parallel to the $z$-axis contribute in (5). Consequently, the induced voltage for a one turn coil becomes [30]:

$$\bar{V} = j\omega l\big(\bar{A}_{z,\mathbf{p}} - \bar{A}_{z,\mathbf{q}}\big) \quad (6)$$

where $\bar{A}_{z,\mathbf{p}}$ and $\bar{A}_{z,\mathbf{q}}$ are the MVP of segments parallel to the $z$-axis located at $\mathbf{p}$ and $\mathbf{q}$ in the $x-y$ plane, respectively. The length of the segments is $l$.

When the imaging region is empty, the induced voltage is proportional to the primary magnetic field and called background voltage. When a conductivity distribution is placed in the imaging region, the induced voltage is proportional to the total magnetic field. The difference in voltage between these two states is proportional to the secondary magnetic field [24].

It is worthwhile mentioning that $\bar{V}$ is a complex value. The real and imaginary parts of $\bar{V}$ have been used in conductivity reconstruction in low and high conductivity applications of MIT, respectively [31]–[33]. It is noteworthy that in some low conductivity studies, the imaginary part of an induced voltage ratio [34] or the phase shift of the induced voltage [14] has been used, as well. Yazdanian and Jafari [35] have shown that these forms are equivalent to using the real part of $\bar{V}$.

*B. MIT Inverse Problem*

MIT inverse problem attempts to obtain images of conductivity distribution inside the imaging region. A variety of algorithms have been developed to solve MIT inverse problem Here, we use the regularized GN algorithm and the enhance technique presented in [25].

*1) Enhanced technique based on the regularized GN*

The enhanced technique uses complex-valued voltages for conductivity reconstruction and works well for all conductivity value ranges, with a superior performance in the middle conductivity values. In this technique, a solution of the MIT inverse problem is attained by minimizing the least-squares objective function $\phi$ given by [25]:

$$\phi(\boldsymbol{\sigma}) = \arg\min_{\boldsymbol{\sigma}} \left\{ \frac{1}{2} \big(\bar{\mathbf{V}}_M - \bar{\mathbf{V}}(\boldsymbol{\sigma})\big)^{H} \big(\bar{\mathbf{V}}_M - \bar{\mathbf{V}}(\boldsymbol{\sigma})\big) + \frac{1}{2} \lambda \boldsymbol{\sigma}^T \mathbf{R}^T \mathbf{R} \boldsymbol{\sigma} \right\} \quad (7)$$

where the superscript H stands for conjugate transpose, $\boldsymbol{\sigma} \in \mathcal{R}^n$ is the conductivity column matrix, $\bar{\mathbf{V}}_M \in \mathcal{C}^m$ denotes the complex-valued column matrix which contains the measured voltages, $\bar{\mathbf{V}}(\boldsymbol{\sigma}): \mathcal{C}^n \to \mathcal{C}^m$ is the complex-valued column matrix obtained from the forward solver, and the numbers $n$ and $m$ represent the number of image pixels and independent measurements, respectively.

Due to the diffuse nature of induced eddy currents, MIT inverse problem is severely ill-posed. Therefore, the last term in right hand side of (7) is added to regularize the problem according to the Tikhonov regularization method. The matrix $\mathbf{R} \in \mathcal{R}^{m \times n}$ is a regularization matrix and $\lambda$ is a regularization parameter.

By the Gauss-Newton optimization of (7) with respect to $\boldsymbol{\sigma}$, one can obtain:

$$\boldsymbol{\sigma}_{k+1} = \boldsymbol{\sigma}_k + \big[\mathbf{J}_k^T \mathbf{J}_k + \lambda_k \mathbf{R}^T \mathbf{R}\big]^{-1} \big[\mathbf{J}_k^T (\mathbf{V}_M - \mathbf{V}_k) - \lambda_k \mathbf{R}^T \mathbf{R} \boldsymbol{\sigma}_k\big] \quad (8)$$

where $\boldsymbol{\sigma}_k$ and $\lambda_k$ are the reconstructed conductivity column matrix and regularization parameter at $k$-th iteration, respectively, $\mathbf{V}_M = [Re\{\bar{\mathbf{V}}_M\} \ Im\{\bar{\mathbf{V}}_M\}]^T$, $\mathbf{V}_k = [Re\{\bar{\mathbf{V}}_k\} \ Im\{\bar{\mathbf{V}}_k\}]^T$, and $\mathbf{J}_k \in \mathcal{R}^{2m \times n}$ is the Jacobian matrix defined by [25]:

$$\mathbf{J}_k = \left[\frac{\partial Re\{\bar{\mathbf{V}}_k\}}{\partial \boldsymbol{\sigma}_k} \ \frac{\partial Im\{\bar{\mathbf{V}}_k\}}{\partial \boldsymbol{\sigma}_k}\right]^T = \frac{\partial \mathbf{V}_k}{\partial \boldsymbol{\sigma}_k}. \quad (9)$$

*2) Choosing regularization parameter*

To select the regularization parameter in (8), we use an adaptive method presented in [20]. This method is automatic and low time consuming. In this method, to obtain an estimate of the Jacobian matrix $\mathbf{J}_h$, a given conductivity value ($\sigma_h$) is first used and a homogeneous conductivity distribution is considered for the imaging region based on $\sigma_h$. Then, to find an initial regularization parameter, the maximum of diagonal elements of $\mathbf{J}_h \mathbf{J}_h^T$ is obtained and multiplied by a coefficient $\tau$. The coefficient is chosen by the user so that the algorithm becomes stable. A single-step Tikhonov method is then applied to obtain an initial conductivity distribution [20]. Finally, the regularization parameter $\lambda_k$ and conductivity column matrix $\boldsymbol{\sigma}_k$ are updated at each iteration.

*C. Jacobian matrix calculation*

The Jacobian matrix, as is computed by the forward problem, is employed to solve the inverse problem. Elements of the matrix specify the sensitivity of simulated voltages to the conductivity of image pixels. The Jacobian matrix is generally calculated using two different techniques in electrical tomography [36]: the sensitivity technique and the standard technique. The former is based on Geselowitz reciprocity theorem [37] and widely used in MIT studies [8], [31], [38]. This method is mainly an independent formulation regardless of the method used in solving the forward problem. In the latter, the Jacobian is directly calculated by the rigorous numerical differentiation of the discretized governing equation in the forward problem, with respect to the electrical conductivity. This method has widely been used in EIT studies with complete electrode model [39], [40], whereas it has only been used in one MIT study [41]. In [41], the standard technique was called the direct method.

Since the sensitivity technique assumes that TCD is a constant space function in coil regions, it cannot be applied to the improved forward method. As a result, we use the standard technique to calculate the Jacobian matrix for the improved method.

*1) Sensitivity technique*

Suppose the phasor voltage $\bar{V}_{ij}$ corresponds to the induced voltage in the $i$-th coil when the $j$-th coil is excited with the current $I_0$ and the sensing coil is excited with unit current. Then, the sensitivity of the simulated voltage $\bar{V}_{ij}$ to the conductivity distribution $\sigma$ is given as follows [25]:

$$\frac{\partial \bar{V}_{ij}}{\partial \sigma} = -\frac{\omega^2}{I_0} \int_\Omega \bar{A}_{z,i} \bar{A}_{z,j} d\Omega \quad (10)$$

where $\Omega$ is the imaging region and $\bar{A}_{z,i}$ and $\bar{A}_{z,j}$ are the solutions of the forward problem (1). If the first order triangular elements are used in FE method, $\bar{A}_z$ inside each element can be approximated as $\bar{A}_z(x,y) \cong \mathcal{N}_e(x,y) \bar{\mathcal{A}}_e$ where $\bar{\mathcal{A}}_e = [\bar{A}_1 \ \bar{A}_2 \ \bar{A}_3]^T$ contains the nodes' potential of the element and $\mathcal{N}_e$ is a matrix containing corresponding shape functions. Then, for each image pixel with the conductivity $\sigma_e$, (10) in discrete form is:

$$\frac{\partial \bar{V}_{ij}}{\partial \sigma_e} = -\frac{\omega^2}{I_0} \bar{\mathcal{A}}_{e,i} \left( \int_{\Omega_e} \mathcal{N}_e \mathcal{N}_e^T d\Omega \right) \bar{\mathcal{A}}_{e,j}^T \quad (11)$$

where $\Omega_e$ is the cross-sectional region of the $e$-th image pixel. One can benefit from the FE method to evaluate the integral in (11) as follows [42]:

$$\frac{\partial \bar{V}_{ij}}{\partial \sigma_e} = -\frac{\omega^2}{I_0} \bar{\mathcal{A}}_{e,i} \mathcal{M}_e \bar{\mathcal{A}}_{e,j}^T \quad (12)$$

The matrix $\mathcal{M}_e$ is defined as follows:

$$\mathcal{M}_e = \begin{bmatrix} \frac{1}{6} & \frac{1}{12} & \frac{1}{12} \\ \frac{1}{12} & \frac{1}{6} & \frac{1}{12} \\ \frac{1}{12} & \frac{1}{12} & \frac{1}{6} \end{bmatrix} \Delta_e \quad (13)$$

where $\Delta_e$ is the area of the $e$-th image pixel.

*2) Standard technique*

As seen from (6), $\bar{V}$ is a function of $\bar{A}_z$. Thus, to obtain $\partial \bar{V}/\partial \sigma$, the calculation of $\partial \bar{A}_z/\partial \sigma$ is required. To get this term in the discrete form, we start with FE equations of the forward problem. After doing FE matrix assembly procedures for (2) or (4), one can find a system of equations as follows:

$$\bar{\mathcal{K}} \bar{\mathcal{A}} = \bar{\mathcal{F}} \quad (14)$$

where $\bar{\mathcal{K}}_{N\times N}$ and $\bar{\mathcal{F}}_{N\times 1}$ are obtained by assembling the left and right hand side of (2) and (4) for the early and improved forward method, respectively, $\bar{\mathcal{A}}_{N\times 1}$ contains the phasors of all node potentials, and $N$ is the total number of FE nodes.

By derivation of (14) with respect to $\sigma_e$, one can obtain:

$$\bar{\mathcal{K}} \frac{\partial \bar{\mathcal{A}}}{\partial \sigma_e} + \frac{\partial \bar{\mathcal{K}}}{\partial \sigma_e} \bar{\mathcal{A}} = 0 \quad (15)$$

or

$$\bar{\mathcal{K}} \frac{\partial \bar{\mathcal{A}}}{\partial \sigma_e} = -\frac{\partial \bar{\mathcal{K}}}{\partial \sigma_e} \bar{\mathcal{A}}. \quad (16)$$

The term $\partial \bar{\mathcal{F}}/\partial \sigma_e$ is zero as the source currents is not dependent upon $\sigma_e$. The Gaussian elimination method can be used to solve the resulting linear system of equations (16) for $\partial \bar{\mathcal{A}}/\partial \sigma_e$. In (16), $\bar{\mathcal{K}}$ and $\bar{\mathcal{A}}$ are known from solving the forward problem. In our application, each element in FEM is a triangle, therefore each element has three nodes. In this case the matrix $\partial \bar{\mathcal{K}}/\partial \sigma_e$ in (16) has at most 9 nonzero elements, no matter how large the dimension of $\bar{\mathcal{K}}$ is. It means this matrix is very sparse.

By considering the discrete form of (6) and derivating with respect to $\sigma_e$, the Jacobian matrix elements in the standard technique can be written as follows:

$$\frac{\partial \bar{V}_{ij}}{\partial \sigma_e} = j\omega l \left( \left[ \frac{\partial \bar{\mathcal{A}}_{ij}}{\partial \sigma_e} \right]_p - \left[ \frac{\partial \bar{\mathcal{A}}_{ij}}{\partial \sigma_e} \right]_q \right) \quad (17)$$

where $\bar{V}_{ij}$ and $\bar{\mathcal{A}}_{ij}$ are the induced voltage and the column matrix containing the phasors of all node potentials when the $j$-th coil is excited and $i$-th coil is measured, respectively, and $\partial \bar{\mathcal{A}}_{ij}/\partial \sigma_e$ is obtained by solving (16). Here, $\left[\partial \bar{\mathcal{A}}_{ij}/\partial \sigma_e\right]_p$ and $\left[\partial \bar{\mathcal{A}}_{ij}/\partial \sigma_e\right]_q$ indicate the $p$-th and $q$-th element of $\partial \bar{\mathcal{A}}_{ij}/\partial \sigma_e$, corresponding to points **p** and **q** in (6), respectively.

### III. NUMERICAL SIMULATIONS

In this section, a 2D MIT problem is modeled and the results of numerical implementation of the inverse problem using both early and improved forward methods are presented and compared. All simulations were executed on a core i5 2.6 GHz laptop with 8 GB of RAM.

*A. Modelling set-up*

Fig. 1(a) shows the cross-sectional view of the 2D MIT system, including eight air-core coils used for both excitation and sensing. The coils are rectangular-shaped and arranged in a circular ring surrounding the imaging region. The imaging region has a radius of 7 cm. Sequential activation of coils using a sinusoidal alternating current of 1 A amplitude excites the imaging region. The previous studies on the low conductivity MIT used excitation frequency in the range of 0.1-13 MHz [7], [43], [44]. Accordingly, we chose 10 MHz for our simulations.

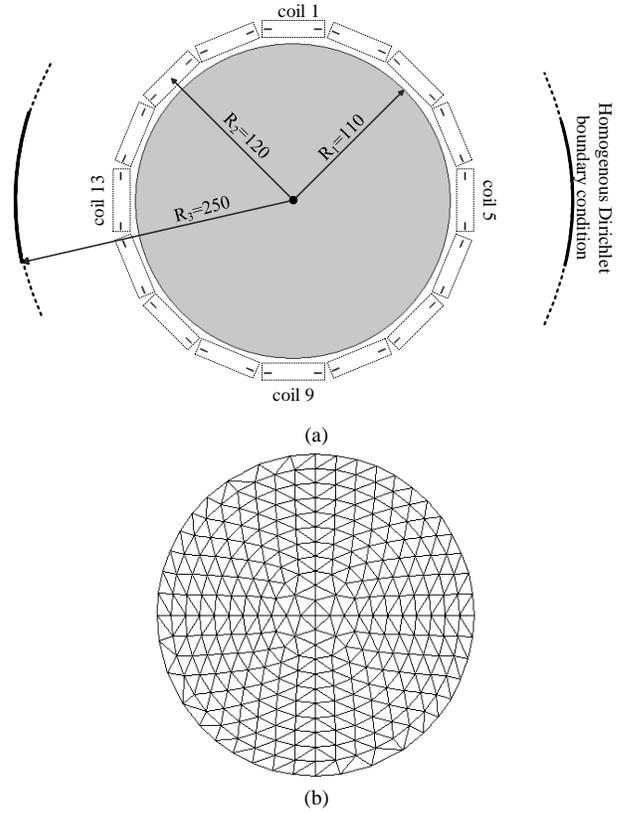

Fig. 1. (a) Coils arrangement and cross section view of the MIT problem model. The homogeneous Dirichlet boundary condition is imposed on a full circle with the radius of 250 mm (the boundary circle was partially drawn for saving space). Dimension is in mm. (b) The mesh including 546 triangular pixels used for solving the inverse problem.

Using higher frequencies is desirable but not allowed because the magneto-quasi-static approximation is no longer valid and it results in more dominant wave-propagation phase delay [45]. After each excitation coil was activated, the induced voltages in the remaining coils (except those previously used for excitation) were measured. The homogeneous Dirichlet boundary condition was imposed at a radius of 10 cm.

In order to evaluate and compare the reconstruction results using the early and improved forward methods, two examples have been considered. In Example I, a circular inclusion with the conductivity of $\sigma_t = 10$ S/m is placed in a background with the conductivity of $\sigma_b = 2$ S/m. In Example I.A, as shown in Fig. 2(a), the inclusion with a radius of 1.5 cm was centered at (-5, 0) cm and in Example I.B, as shown in Fig. 3(a), the inclusion with a radius of 1 cm was centered at (-5.5, 0) cm. In Example II, two circular inclusions with the conductivity of $\sigma_{t1} = 5$ S/m and $\sigma_{t2} = 10$ S/m are placed in a background with the conductivity of $\sigma_b = 2$ S/m. In Example II.A, as shown in Fig. 4(a), $\sigma_{t1}$ and $\sigma_{t2}$ with a radius of 1.5 cm were centered at (-3.5, 3.5) cm and (-3.5, -3.5) cm, respectively. In Example II.B, as shown in Fig. 5(a), $\sigma_{t1}$ and $\sigma_{t2}$ with a radius of 1 cm were centered at (-3.9, 3.9) cm and (-3.9, -3.9) cm, respectively. The theoretical limit given for the minimum detectable inhomogeneity radius for the modelled system is $r_{min}= 1.3$ cm. This limit is obtained from $r_{min} = R/\sqrt{m}$, where $R$ and $m$ are the radius of imaging region and the number of independent measurements, respectively [1]. Since the radius 1 cm of the small target object is a little smaller than the limit, we placed the center of the small target object closer to the boundary.

The forward problem is solved by the FE method based on the early and improved forward methods. The overall number of triangular elements and nodes in the FE model was 768 and 409, respectively. The inverse problem was solved by the GN algorithm based on the technique presented in [25]. As shown in Fig. 1(b), the mesh including 294 uniform triangular pixels was used to solve the inverse problem. As illustrated, pixels have almost the same size. In addition, to avoid an inverse crime, the simulated measured data has been produced by solving the improved forward method on a very fine mesh with about $10^5$ triangular elements and $5 \times 10^4$ nodes.

### B. Performance parameters

To evaluate the reconstructed images in Example I, we use four performance parameters (PPs): conductivity contrast (CC), resolution (RES), position error (PE), and relative error (RE). To define CC, RES, and PE, a threshold is applied to the reconstructed image as follows:

$$[\boldsymbol{\sigma}^t]_i = \begin{cases} 1 & if \ [\boldsymbol{\sigma}]_i > \sigma_{thr} \\ 0 & \text{otherwise.} \end{cases} \quad (18)$$

where $[\boldsymbol{\sigma}]_i$ and $[\boldsymbol{\sigma}^t]_i$ are the $i$-th image pixel and the $i$-th thresholded amplitude image pixel, respectively. In the binary column matrix $\boldsymbol{\sigma}^t$, the non-zero elements correspond to image pixels whose conductivity value exceeds the threshold $\sigma_{thr}$. The threshold value provides a trade-off to distinguish between the visually important effects and background in the reconstructed image.

*Conductivity contrast (CC)* measures the ratio between the conductivity of the reconstructed target object to that of its surrounding background [46]. The target object and background conductivity values are determined based on thresholded amplitude set of the reconstructed image. Then, the average of pixels' conductivities labeled as the background ($\sigma_b$) and target object ($\sigma_t$) are calculated and CC is obtained as $\sigma_t/\sigma_b$.

*Resolution (RES)* is calculated as [47]:

$$\text{RES} = \sqrt{A^t/A^0} \quad (19)$$

where $A^t = \sum_k [\boldsymbol{\sigma}^t]_k$ is the number of pixels greater than $\sigma_{thr}$ and $A^0$ is the area (in pixels) of the entire imaging region.

*Position error (PE)* shows the position discrepancy between the centroid of the target object in the reconstructed image and the simulated medium. PE is defined by [47]:

$$\text{PE} = r_t - r_h \quad (20)$$

where $r_t$ and $r_h$ are the radial position of the centroid of the actual target and reconstructed target, respectively. It is desired that PE is small and shows low variability for targets at different radial positions.

*Relative error* for reconstructed conductivity image at $k$-th iteration is calculated as:

$$\text{RE}_k(\%) = \frac{\|\boldsymbol{\sigma}^{true} - \boldsymbol{\sigma}_k\|_2}{\|\boldsymbol{\sigma}^{true}\|_2} \quad (21)$$

where $\|\cdot\|_2$ denotes $L^2$ norm and $\boldsymbol{\sigma}^{true}$ is a column matrix contains true conductivity distribution.

### C. Example I: One target object

In this example, the imaging region includes one target object with a radius of 1.5 cm in Example I.A and with a radius of 1 cm in Example I.B. Fig. 2(b)-(d) illustrate the reconstructed conductivity images by using the early forward problem for Example I.A. The homogeneous conductivity value $\sigma_h$ and the coefficient $\tau$ were 1 S/m and 3, respectively. The voltages induced by the secondary and total magnetic fields based on the early forward method are indicated by $\Delta \mathbf{V}^E$ and $\mathbf{V}^E$, respectively. In Fig. 2(b) and Fig. 2(c), $\Delta \mathbf{V}^E$ have been used for reconstruction and two different colorbar scales have been applied to display the results. In Fig. 2(b), the colorbar is scaled to the minimum and maximum value of estimated conductivity values and, in Fig. 2(c); it is scaled to [0  10] S/m. As can be seen, using the voltages induced by the secondary field can partially compensate for the impact of ignoring skin and proximity effects in the early forward method while it sacrifices the conductivity contrast in the reconstructed image. In Fig. 2(d), $\mathbf{V}^E$ has been used for reconstruction. As can be seen, when the voltages induced by the total magnetic field are computed by the early forward method, conductivity distribution is not meaningfully reconstructed. It means that ignoring skin and proximity effects in coils in the forward problem implicates considerable errors in the reconstructed image, as explained in [24].

Fig. 2(e)-(g) illustrate the reconstructed conductivity images by using the improved forward problem for Example I.A. The homogeneous conductivity value $\sigma_h$ and the coefficient $\tau$ were 1 S/m and 3, respectively. The voltages induced by the

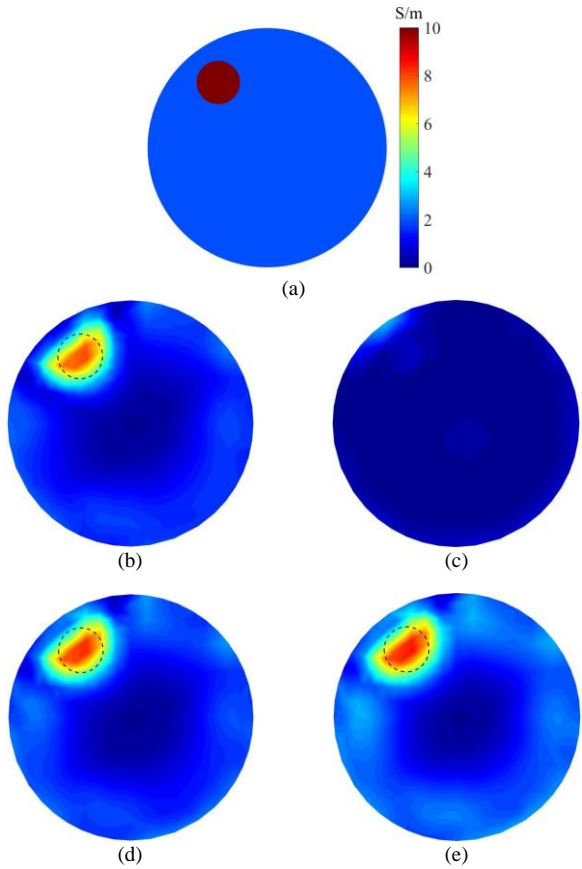

Fig. 2. Example I: Imaging region contains one target object with radius of 20 mm. (a) True conductivity distribution. Reconstructed conductivity images using (b) the early forward method and the secondary field, (c) the early forward method and the total field, (d) the improved forward method and the secondary field, and (e) the improved forward method and the total field. The homogeneous conductivity value $\sigma_h$ the coefficient $\tau$ were 0.5 S/m and 3.5, respectively. The target object and background conductivity were $\sigma_t = 10$ S/m and $\sigma_b = 2$ S/m, respectively.

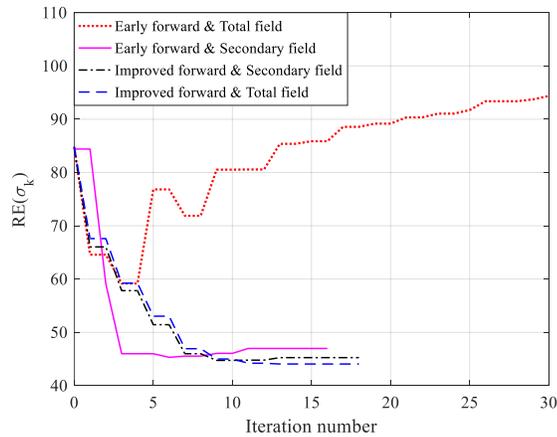

Fig. 3. Example I: Relative error versus iteration number for each case.

TABLE I
EXAMPLE I.A: PERFORMANCE PARAMETERS (PPs) COMPUTED FOR DIFFERENT CASES OF FORWARD METHOD AND MAGNETIC FIELD USED FOR COMPUTATION OF INDUCED VOLTAGE. CASE I: EARLY FORWARD METHOD AND SECONDARY MAGNETIC FIELD, CASE II: IMPROVED FORWARD METHOD AND SECONDARY MAGNETIC FIELD, AND CASE III: IMPROVED FORWARD METHOD AND TOTAL FIELD. THE PARAMETER $K$ INDICATES THE ITERATION NUMBER FOR EACH CASE. PPs ARE EXPLAINED IN TEXT.

| Case | Jacobian technique | PPs | | | | | | |
|---|---|---|---|---|---|---|---|---|
| | | $\sigma_t$ | $\sigma_b$ | CC | RES | PE | $K$ | Time |
| | | S/m $10^*$ | S/m $2^*$ | - $5^*$ | - $0.2^*$ | mm $0^*$ | - - | (min) - |
| Case I | Standard | 7.5 | 1.3 | 5.8 | 0.2 | -1 | 16 | 1.1 |
| Case I | Sensitivity | 7.5 | 1.3 | 5.8 | 0.2 | -1 | 16 | 6.9 |
| Case II | Standard | 7.8 | 1.4 | 5.7 | 0.2 | -0.3 | 18 | 7.8 |
| Case III | Standard | 7.9 | 1.7 | 4.6 | 0.2 | -0.1 | 18 | 7.8 |

* Ideal value

improved forward method is applied, using voltages induced by both total and secondary magnetic fields can detect the target object. However, it seems that using $\mathbf{V}^I$ results in a visually better reconstructed image.

Table I indicates PPs obtained for Example I.A. Since using $\mathbf{V}^E$ in inverse problem could not meaningfully reconstruct the conductivity distribution, PPs are indeterminable. Thus, they are not reported in Table I. The parameter $K$ indicates the secondary and total magnetic fields based on the improved forward method are indicated by $\Delta\mathbf{V}^I$ and $\mathbf{V}^I$, respectively.

In Fig. 2(e) and Fig. 2(f), $\Delta\mathbf{V}^I$ have been used for reconstruction and two different colorbar scales have been applied to display the results. In Fig. 2(e), the colorbar is scaled to the minimum and maximum value of estimated conductivity values and, in Fig. 2(f); it is scaled to [0 10] S/m. In Fig. 2(g), $\mathbf{V}^I$ has been used for reconstruction. As can be seen, when the iteration number for each case. As can be seen, using $\mathbf{V}^I$ results in the best performance except for CC. For the early forward method, we tested both sensitivity and standard techniques for Jacobian matrix calculation. The reconstructed images were the same. However, the runtime was different. As expected, the standard technique was more time-consuming.

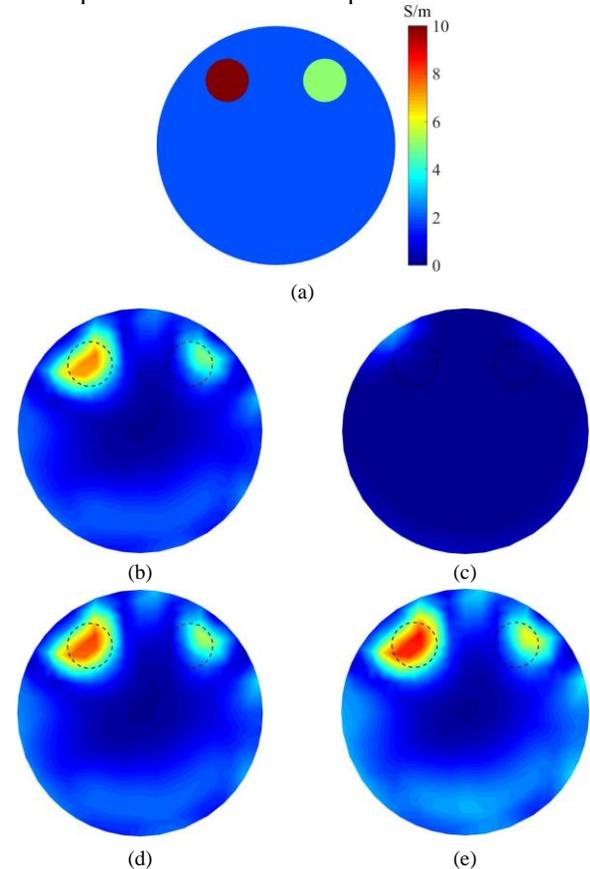

Fig. 4. Example II: Imaging region contains two target objects with radius of 20 mm. (a) True conductivity distribution. Reconstructed conductivity images using (b) the early forward method and the secondary field, (c) the early forward method and the total field, (d) the improved forward method and the secondary field, and (e) the improved forward method and the total field. The homogeneous conductivity value $\sigma_h$ and the coefficient $\tau$ were 0.5 S/m and 3.5, respectively. The target object conductivities were $\sigma_{t1} = 10$ S/m (left target) and $\sigma_{t2} = 5$ S/m (right target) and the background conductivity was $\sigma_b = 2$ S/m.

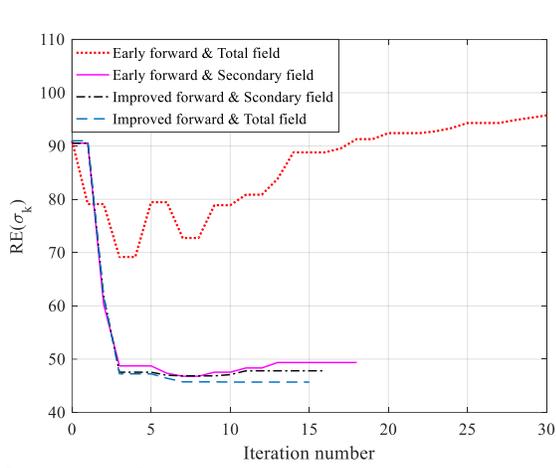

Fig. 5. Example II: Relative error versus iteration number for each case.

Fig. 3(b)-(d) illustrate the reconstructed conductivity images by using the early forward problem for Example I.B. The homogeneous conductivity value $\sigma_h$ and the coefficient $\tau$ were 1 S/m and 4, respectively. In Fig. 3(b) and Fig. 3(c), $\Delta \mathbf{V}^E$ have been used for reconstruction and two different colorbar scales have been applied to display the results. In Fig. 3(b), the colorbar is scaled to the minimum and maximum value of estimated conductivity values and, in Fig. 3(c); it is scaled to [0 10] S/m. Similar to Example I.A, using the voltages induced by the secondary field can partially compensate for the impact of ignoring skin and proximity effects in the early forward method while it sacrifices the conductivity contrast in the reconstructed image. However, in Fig. 3(c) compared to Fig. 2(c), the target object was barely detected. It means that when the target object becomes smaller, the compensatory effect of using the secondary field becomes less. In Fig. 3(d), $\mathbf{V}^E$ has been used for reconstruction. As can be seen, when the voltages induced by the total magnetic field are computed by the early forward method, conductivity distribution is not meaningfully reconstructed.

Fig. 3(e)-(g) illustrate the reconstructed conductivity images by using the improved forward problem for Example I.B. The homogeneous conductivity value $\sigma_h$ was 1 S/m. The coefficient $\tau$ was 3 and 2.2 when the voltages induced by the secondary and total fields are used, respectively. In Fig. 3(e) and Fig. 3(f), $\Delta \mathbf{V}^E$ have been used for reconstruction and two different colorbar scales have been applied to display the results. In Fig. 3(e), the colorbar is scaled to the minimum and maximum value of estimated conductivity values and, in Fig. 3(f); it is scaled to [0 10] S/m. In Fig. 3(g), $\mathbf{V}^I$ has been used for reconstruction. Similar to Example I.A, when the improved forward method is applied, using voltages induced by both total and secondary magnetic fields can detect the target object.

Table II indicates PPs obtained for Example I.B. Similar to Example I.A, PPs are not reported for Fig. 3(d). As can be seen, using $\mathbf{V}^I$ results in the best performance except for PE. As seen from Fig 3(g), the center of reconstructed target object slightly moved towards the origin. Table II shows the reconstructed $\sigma_t$ and $\sigma_b$ were 2 and 0.89, respectively, when $\Delta \mathbf{V}^E$ was used, respectively. Consequently, in Fig. 3(c), the target object cannot be distinguished according to the given colorbar scale. Similar to Example I.A, the standard technique was more time-consuming.

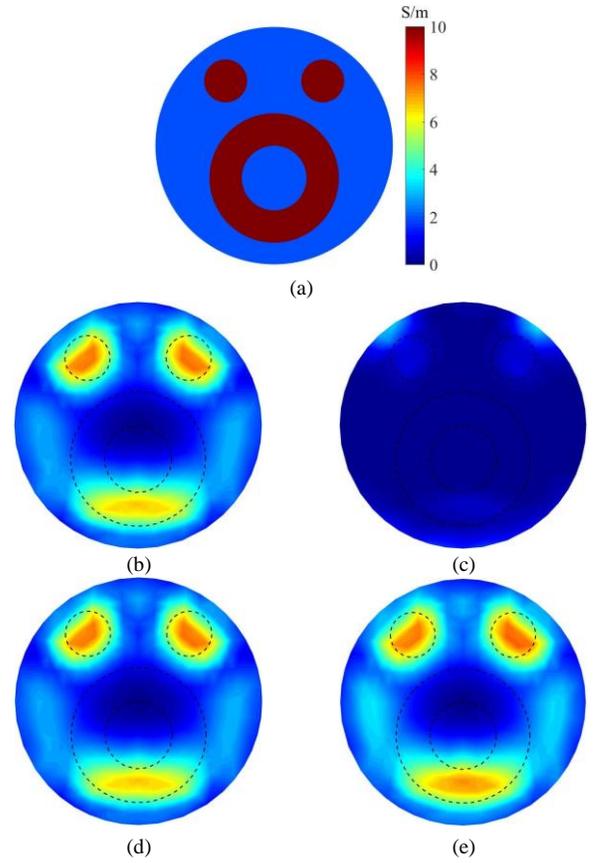

Fig. 6. Example III: Imaging region contains two target objects with radius of 20 mm and a ring with thickness of 30 mm. (a) True conductivity distribution. Reconstructed conductivity images using (b) the early forward method and the secondary field, (c) the early forward method and the total field, (d) the improved forward method and the secondary field, and (e) the improved forward method and the total field. The homogeneous conductivity value $\sigma_h$ and the coefficient $\tau$ were 0.5 S/m and 4, respectively. The target object conductivities were 10 S/m and the background conductivity was $\sigma_b = 2$ S/m.

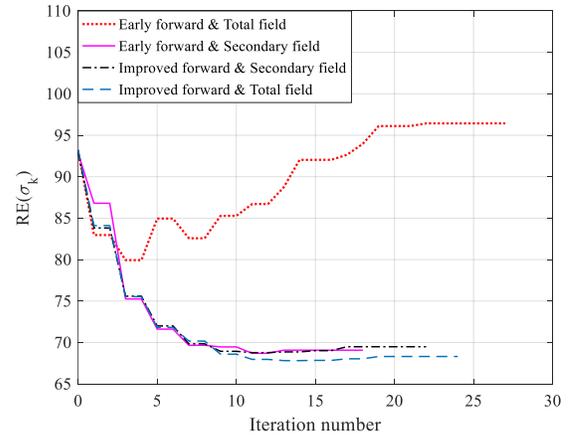

Fig. 7. Example II: Relative error versus iteration number for each case.

*D. Example II: Two target objects*

In this example, the imaging region includes two target objects with the target conductivities $\sigma_{t1} = 5$ S/m (upper target) and $\sigma_{t2} = 10$ S/m (lower target). The radius of targets is 1.5 cm in Example II.A and 1 cm in Example II.B.

Fig. 4 illustrates the reconstructed conductivity images for Example II.A. The homogeneous conductivity value $\sigma_h$ and the coefficient $\tau$ were 1 S/m and 3, respectively. In Fig. 4(b)-(d), $\Delta \mathbf{V}^E$, $\Delta \mathbf{V}^I$, and $\mathbf{V}^I$ have been used for reconstruction, respectively. Similar to the previous examples, when the voltages induced by the total magnetic field are computed by the early forward method, conductivity distribution is not meaningfully reconstructed and the corresponding image is not shown in Fig. 4. As shown in Fig. 4(b), using $\Delta \mathbf{V}^E$ in this example partially detects the target object with higher conductivity value and the target with lower conductivity cannot be distinguished. When the improved forward method is applied, using voltages induced by both total and secondary magnetic fields can detect both target objects. However, using $\Delta \mathbf{V}^I$ results in lower contrast in the reconstructed images. The percentage of relative error at the final iteration was $RE_{30}=67\%$, $RE_{30}=35\%$, and $RE_{27}=32\%$ when $\Delta \mathbf{V}^E$, $\Delta \mathbf{V}^I$, and $\mathbf{V}^I$ were used, respectively.

Fig. 5 illustrates the reconstructed conductivity images for Example II.B. The homogeneous conductivity value $\sigma_h$ was 1 S/m. The coefficient $\tau$ was 4 and 3 for the early and improved forward methods, respectively. In Fig. 5(b)-(d), $\Delta \mathbf{V}^E$, $\Delta \mathbf{V}^I$, and $\mathbf{V}^I$ have been used for reconstruction, respectively. Similar to the previous examples, when the voltages induced by the total magnetic field are computed by the early forward method, conductivity distribution is not meaningfully reconstructed and the related image is not shown in Fig. 5. As shown in Fig. 5(b), when $\Delta \mathbf{V}^E$ is used the target objects cannot be distinguished in the [0  10] colorbar scale. When the improved forward method is applied, using voltages induced by both total and secondary magnetic fields can detect both target objects. However, using $\Delta \mathbf{V}^I$ results in lower contrast in the reconstructed images. The percentage of relative error at the final iteration was $RE_{22}=67\%$, $RE_{26}=38\%$, and $RE_{19}=36\%$ when $\Delta \mathbf{V}^E$, $\Delta \mathbf{V}^I$, and $\mathbf{V}^I$ were used, respectively.

*E. Noise study*

In this subsection, we study the robustness of the reconstruction algorithm against the noise when $\Delta \mathbf{V}^E$, $\Delta \mathbf{V}^I$, and $\mathbf{V}^I$ are used. For this purpose, we chose Example I.A in which the target object was detected when $\Delta \mathbf{V}^E$, $\Delta \mathbf{V}^I$, and $\mathbf{V}^I$ were used. We added complex white Gaussian noise to the simulated measured voltages and considered 40, 30, and 20 dB signal to noise ratio (SNR). For each SNR, we repeated the experiment 50 times. To evaluate the performance of the reconstruction, we used the thresholded amplitude conductivity image obtained by (18). For SNR= 40 dB, the target object was detected in all 50 thresholded images when $\Delta \mathbf{V}^E$, $\Delta \mathbf{V}^I$, or $\mathbf{V}^I$ was used. Furthermore, the average RE was $69 \pm 0.3\%$, $46 \pm 0.4\%$, and $40 \pm 1.2\%$ when $\Delta \mathbf{V}^E$, $\Delta \mathbf{V}^I$, and $\mathbf{V}^I$ were used, respectively. When SNR decreased to 30 dB, using $\Delta \mathbf{V}^E$ in GN algorithm resulted in detection of target objects in 42 thresholded images (out of 50), using $\Delta \mathbf{V}^I$ in GN algorithm resulted in detection of target objects in 49 thresholded images (out of 50) and using $\mathbf{V}^I$ resulted in detection of target objects in all thresholded images. Furthermore, the average RE was $70 \pm 1.1\%$, $47 \pm 1.8\%$, and $41 \pm 2.3\%$ when $\Delta \mathbf{V}^E$, $\Delta \mathbf{V}^I$, and $\mathbf{V}^I$ were used, respectively. By decreasing the SNR to 20 dB, using $\Delta \mathbf{V}^E$, $\Delta \mathbf{V}^I$, and $\mathbf{V}^I$ resulted in detection of target objects in 24, 29, and 38 thresholded images, respectively. Furthermore, the average RE was $101 \pm 82\%$, $65 \pm 38\%$, and $48 \pm 7.1\%$ when $\Delta \mathbf{V}^E$, $\Delta \mathbf{V}^I$, and $\mathbf{V}^I$ were used, respectively.

IV. DISCUSSION

As seen in Example I and Example II, when $\mathbf{V}^E$, induced voltages obtained from the total field and computed by the early forward method, is used in the inverse problem, the conductivity distribution is not reconstructed meaningfully. It manifests that ignoring skin and proximity effects inside MIT coils in the forward problem implicates considerable errors in the reconstructed image. As mentioned in [24], using gradiometer or state-difference imaging techniques to obtain voltages induced by the secondary magnetic field, $\Delta \mathbf{V}^E$, can partially compensate the error due to neglecting of skin and proximity effects in coils. However, as seen in Examples I.B and II.B, when target objects become smaller, it is hard to distinguish them in the reconstructed images. In addition, in Example II.A where large target objects are placed in the imaging region, using $\Delta \mathbf{V}^E$ partially reconstructs the target object with higher conductivity and the target with lower conductivity remains unresolvable.

It is noteworthy that the reconstructed conductivity values using the voltages induced by the secondary magnetic field, $\Delta \mathbf{V}^E$ and $\Delta \mathbf{V}^I$, are lower compared to the true conductivities. In other words, using the secondary magnetic field data causes the conductivity values to be underestimated. Consequently, using the secondary magnetic field data to compensate for error due to neglecting skin and proximity effects in coils, will be at the cost of producing qualitative images. As seen in Section III, to reconstruct the absolute conductivity values, it is necessary to use the total field data and considering skin and proximity effects inside MIT coils.

Here, we are not dealing with a linear system in which the superposition principle can be applied. Removing the primary field compensates partially for the error caused by ignoring the skin and proximity effects in coils, but not completely. In fact, by placing the target object in the imaging region, losses caused by the skin and proximity effects in coils change compared to when the imaging region is empty (primary field). This change is due to interaction between coils and the conductivities to be imaged as shown in [24]. Consequently, removing the primary field from the total field cannot completely compensate for ignoring the skin and proximity effects in coils. In addition, as seen in Section III.E, using the secondary data has another drawback. The reconstruction procedure based on the secondary field data has less robustness against the noise.

As expected and seen from the simulation results, the standard technique for calculation of the Jacobian matrix is more computationally demanding compared to the sensitivity one. In this work, we observed both standard and sensitivity techniques had the same performance in terms of the reconstructed conductivity for the early forward method.

However, in [41], it has been shown that the standard technique is more accurate in some situations.

## V. CONCLUSION

In this paper, numerical conductivity image reconstruction based on the improved forward method was developed for 2D MIT. Improved forward method is based on complete Maxwell's equations and considers skin and proximity effects inside the exciter and sensor coils. Using improved forward method in the MIT conductivty reconsturction procedure was investigated by modeling an 8-coil 2D MIT system through two different numerical experiments. Results of this study manifested that the error due to neglecting the skin and proximity effects can be partially compensated by the difference imaging; however, it will be at the cost of producing qualitative images. Furthermore, to reconstruct the absolute conductivity values in the low conductivity MIT applications, it is crucial to use the improved forward method and voltages induced by the total magnetic field.